\documentclass[useAMS,usegraphicx,usenatbib]{mn2e}
\usepackage{color} 

\usepackage{multirow}
\usepackage{amssymb}

\usepackage{amsmath}

\newcommand{\kms}{\mbox{$\>{\rm km\, s^{-1}}$}}
\newcommand{\masyr}{\mbox{$\>{\rm mas\, yr^{-1}}$}}
\newcommand{\kpc}{\mbox{$\>{\rm kpc}$}} 
\newcommand{\pc}{\mbox{$\>{\rm pc}$}} 
\newcommand{\Gyr}{\mbox{$\>{\rm Gyr}$}}
\newcommand{\yr}{\mbox{$\>{\rm yr}$}}
\newcommand{\Msun}{\>{\rm M_{\odot}}}
 
\newcommand{\rsun}{\mbox{$R_{\rm s}$}} 
\newcommand{\vgsr}{\mbox{$V_{\rm GSR}$}} 
\newcommand\degrees{^\circ}
\newcommand{\avg}[1]{\mbox{$\left<{#1}\right>$}}

\newcommand{\pmo}[1]{\mbox{$\mu_{#1}$}}
\newcommand{\tm}[1]{\mbox{$t_{#1}$}}

\def\eg{{\it e.g.}}

\def\ie{{\it i.e.}}

\title[Predicted Kinematics of a kpc-Scale Disc (or Ring)]{Predicted Stellar
  Kinematics of a Kiloparsec-Scale Nuclear Disc (or Ring) in the Milky
  Way}
   
\author[Debattista et al.]{Victor P. Debattista$^{1}$\thanks{E-mail:
    vpdebattista@gmail.com}, Samuel W. F. Earp$^1$, Melissa Ness$^2$,
  Oscar A. Gonzalez$^{3}$ \\ $^1$
  Jeremiah Horrocks Institute, University of Central Lancashire,
  Preston, PR1 2HE, UK \\ $^2$ Max-Planck-Institut f\"ur Astronomie,
  K\"onigstuhl 17, D-69117 Heidelberg, Germany \\ $^3$ UK Astronomy
  Technology Centre, Royal Observatory, Blackford Hill, Edinburgh, EH9
  3HJ, UK\\ }

\begin{document}   

\date{{\it Draft version on \today}}
\pagerange{\pageref{firstpage}--\pageref{lastpage}} \pubyear{----}
\maketitle

\label{firstpage}

\begin{abstract} 
  In \citet{debattistaMWND+15}, we proposed that a kiloparsec-scale
  nuclear disc is responsible for the high-velocity secondary peak in
  the stellar line-of-sight velocity distributions (LOSVDs) seen at
  positive longitudes in the bulge by the Apache Point Observatory
  Galactic Evolution Experiment (APOGEE).  Here, we make further
  qualitative but distinctive predictions of the kinematic properties
  of a nuclear disc, including for the LOSVDs at negative longitudes
  (which APOGEE-2 will observe) and examine the proper motions
  throughout the disc. Since a nuclear ring is also able to produce
  similar high-velocity LOSVD peaks, we present predictions for the
  proper motion signatures which distinguish between a nuclear disc
  and a nuclear ring.  We also demonstrate that the stars in a nuclear
  disc, which would be on x2 orbits perpendicular to the bar, can
  remain on these orbits for a long time and can therefore be old.  We
  show that such (old) nuclear discs of comparable size exist in
  external galaxies.
\end{abstract}


%

\section{Introduction}
\label{sec:intro}

Using Apache Point Observatory Galactic Evolution Experiment
\citep[APOGEE][]{alam+15} commissioning data, \citet{nidever+12}
studied the line-of-sight velocity distributions (LOSVDs) of stars
within the central regions of the Milky Way (MW).  They found a
secondary (high) peak in Galactocentric velocity (at $\vgsr \approx
200 \kms$) for fields in and near the plane, which they proposed is
composed of stars on bar orbits.  However the pure $N$-body models of
\citet{li+14} lack such cool, high-\vgsr\ peaks.  \citet{li+14} also
noted a lack of corresponding peaks at the opposite longitudes in the
Bulge Radial Velocity Assay \citep[BRAVA][]{kunder+12} data.  Their MW
$N$-body model instead showed that the LOSVDs have shoulders extending
to large velocities coming from stars at large distance from the Sun.
\citet{gomez+16} find a similar result, also using pure $N$-body
simulations; they fit two Gaussians to the LOSVDs and find that a cool
high-velocity component is needed, but that these do not produce the
trough observed in the LOSVD.  Using the simulation of \citet{li+14},
\citet{molloy+15} showed that resonant orbits produced
high-\vgsr\ peaks.  \citet[][hereafter AS15]{aumer_schoenrich15}
argued that young stars recently trapped by the bar into resonant
orbits are favoured by the selection function of the APOGEE survey.
Based on $N$-body simulations, they proposed that it is preferentially
young stars that give rise to the high-\vgsr\ peaks.
However \citet{zasowski+16} and \citet{zhou_shen+17} recently showed
that stars in the APOGEE high-\vgsr\ peaks do not exhibit distinct
chemical abundances or ages, indicating that they are not
predominantly comprised of younger stars.  As there is no reason for
the bar to have stopped growing in the past few gigayears, other
models for the high-\vgsr\ peaks need to be considered.

Based on comparison with an $N$-body$+$smooth particle hydrodynamics
(SPH) simulation with gas and star formation, \citet[][hereafter
  D15]{debattistaMWND+15} proposed that the high-\vgsr\ peaks in the
mid-plane at $l = 6\degrees-8\degrees$ reveal the presence of a
kiloparsec-scale nuclear disc, supported by x2 orbits aligned
perpendicular to the bar.  In their simulation, the nuclear disc forms
when gas is driven to the centre by the bar and forms stars.
\citet{schoenrich+15} argued that a kiloparsec (kpc) scale is too
large a disc for the MW if it forms out of gas reaching the centre
now, showing instead that a nuclear disc of 150 \pc\ size is present
at the centre of the MW.  The presence of a 150 \pc-sized nuclear disc
in the MW is unsurprising, given that gas now being funnelled by the
bar settles into a ring of about this radius \citep{binney+91,
  weiner_sellwood99, sormani+15, lizhi+16}.  However \citet{cole+14}
showed that nuclear discs of size comparable to that in the model of
D15 exist in early-type galaxies.  Moreover such nuclear discs can be
comprised of old stars \citep[e.g.][]{gadotti+15}, alleviating the
problem of needing young stars in the high-\vgsr\ peak.

In order to help test whether a kpc-sized nuclear disc or ring exists
in the MW, here we present several predictions for the resulting
kinematics, which ongoing (\eg\ APOGEE-2) and future (\eg\ MOONS)
surveys can test.  We use the same simulation as in D15, observed in
the same way, to predict properties of the kinematics that are
characteristic of a nuclear disc, particularly at the negative
longitudes that have not yet been probed by large surveys.  APOGEE-2,
in particular, will, in the next few years, measure velocities for
bulge stars across negative longitudes, including in the plane. This
will provide a homogeneous set of high-resolution spectroscopic data
for about 50,000 stars toward the bulge and inner disc across negative
and positive longitudes and latitudes \citep{apogee}.  We show that
the model provides several clear kinematic diagnostics of a nuclear
disc or ring from such data.


\section{High velocity peaks from old stars on x2 orbits}
\label{sec:ages}

D15 showed that a nuclear disc formed in their simulation after
$6\Gyr$ and was well developed by $7.5\Gyr$, at which point they used
their model to present the LOSVD signatures of a nuclear disc.
Because the nuclear disc formed late in their simulation, of necessity
the nuclear disc they presented was young.  Here we explore whether
the nuclear disc {\it must} be a young structure, or whether it
survives as an old structure.  In order to test this, we evolve the
model of D15 from $7.5\Gyr$ to $13.5\Gyr$ with gas cooling and star
formation turned off.  During this evolution, the bar roughly doubles
in size and the pattern speed drops to $30\%$ of its value at
$t=7.5\Gyr$.  At that point we repeat the same analysis shown in
Figure 1 of D15, presenting the LOSVDs in the mid-plane and at
$b=2\degrees$.  We scale the model exactly as in D15, in order to be
able to compare directly with that earlier work.

\begin{figure}
\centerline{
\includegraphics[angle=0.,width=1.\hsize]{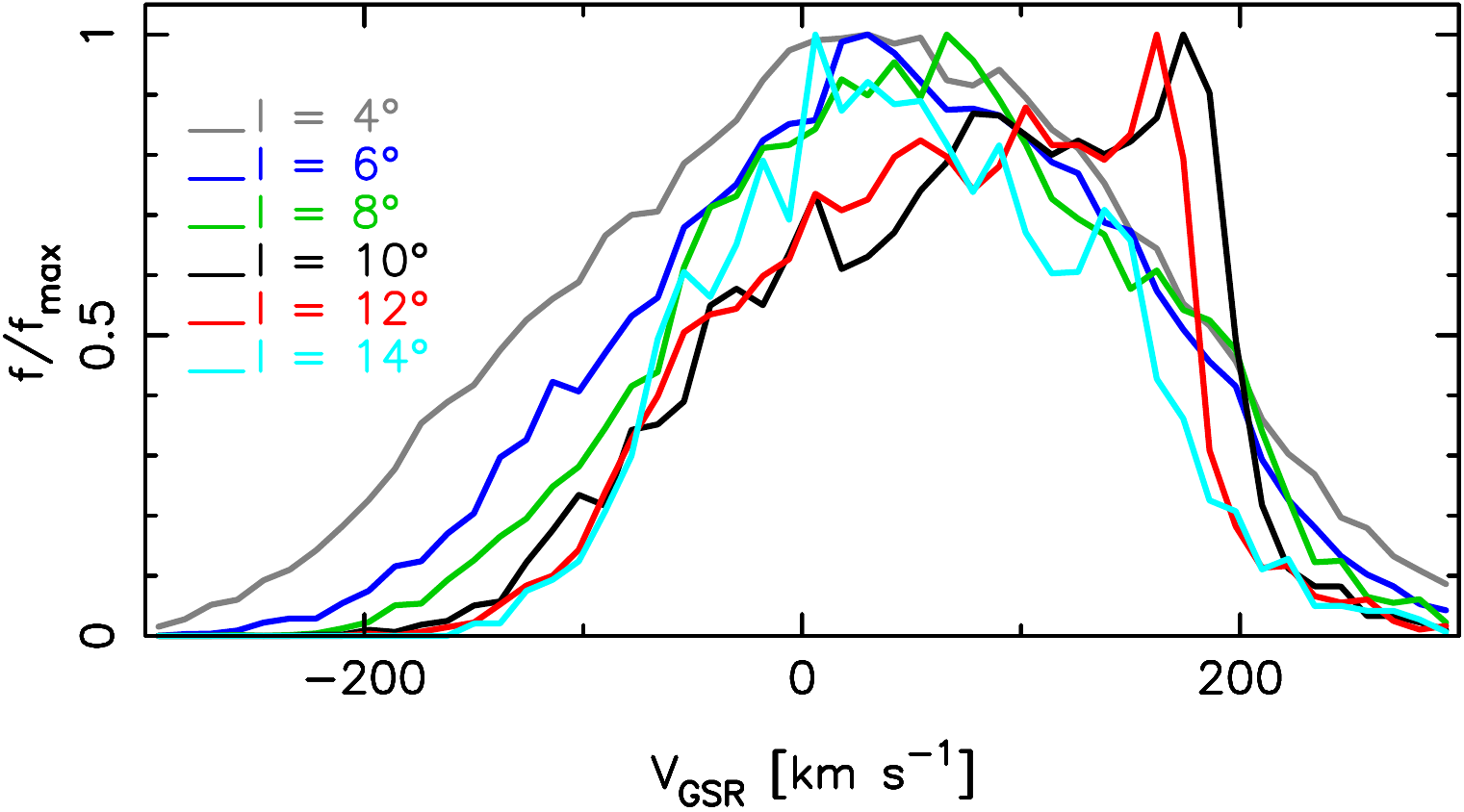}\
}
\centerline{
\includegraphics[angle=0.,width=1.\hsize]{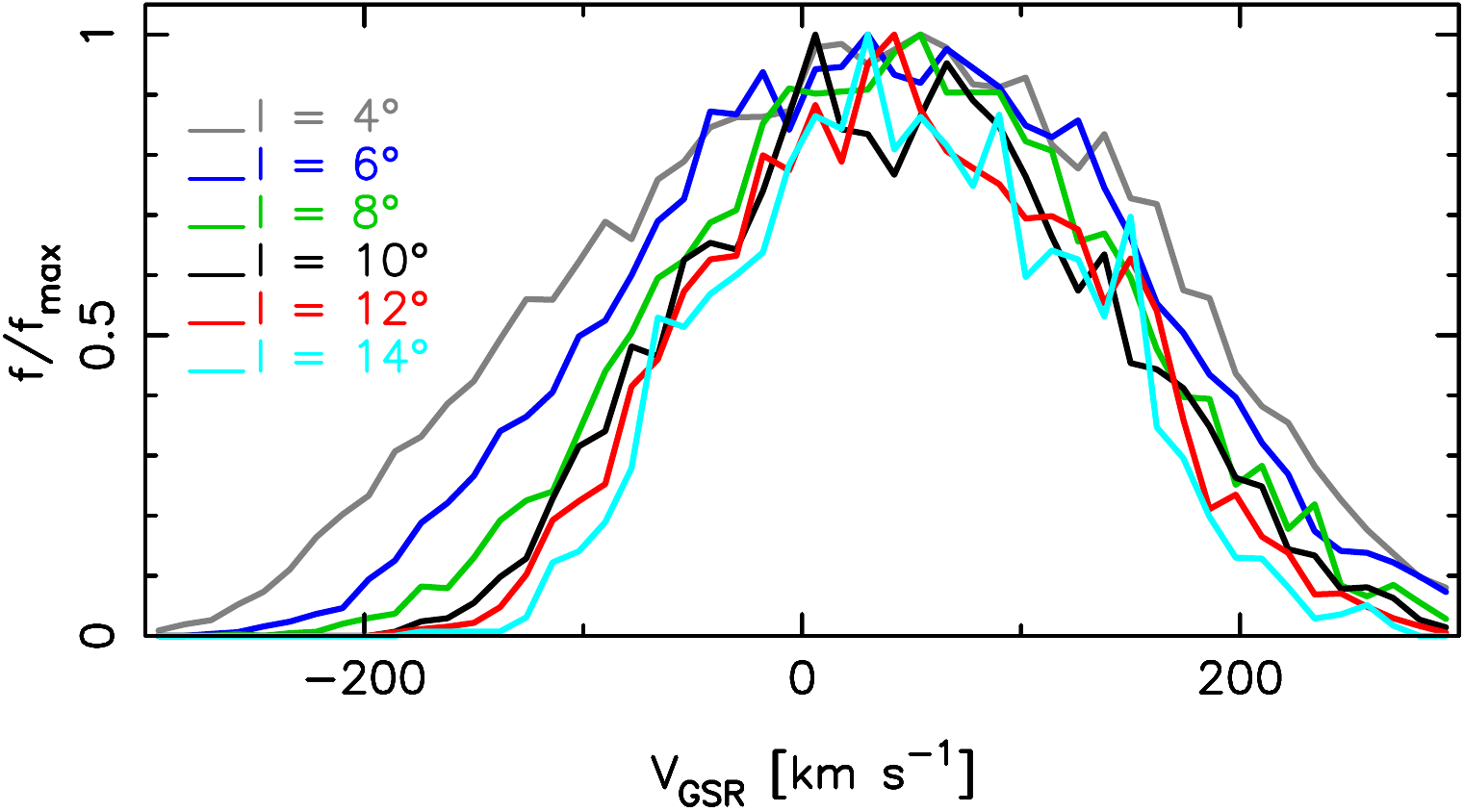}\
}
\caption{LOSVDs in the model of D15 evolved for a further $6 \Gyr$
  with no star formation.  Top: In the mid-plane ($b=0\degrees$).
  Bottom: at $b=2\degrees$.  Compare with the right columns of Figure
  1 of D15.}
\label{fig:olddisk}
\end{figure}

The LOSVDs in the same lines of sight as in D15 are shown in
Fig. \ref{fig:olddisk}.  In spite of the strong evolution of the bar,
high velocity peaks are still evident in the mid-plane (top panel) at
$l= \pm10\degrees$ and $l = \pm 12\degrees$, although the peak at
$l=\pm 8\degrees$ has disappeared.  Just as remarkably, the model
still retains no signature of a high velocity peak at $b=2\degrees$
(bottom panel).  We will explore the evolution of the peaks in more
detail using an orbital analysis elsewhere (Earp et al. in progress).

\subsection{Comparison with external galaxies}

\begin{figure}
\includegraphics[angle=0.,width=0.9\hsize]{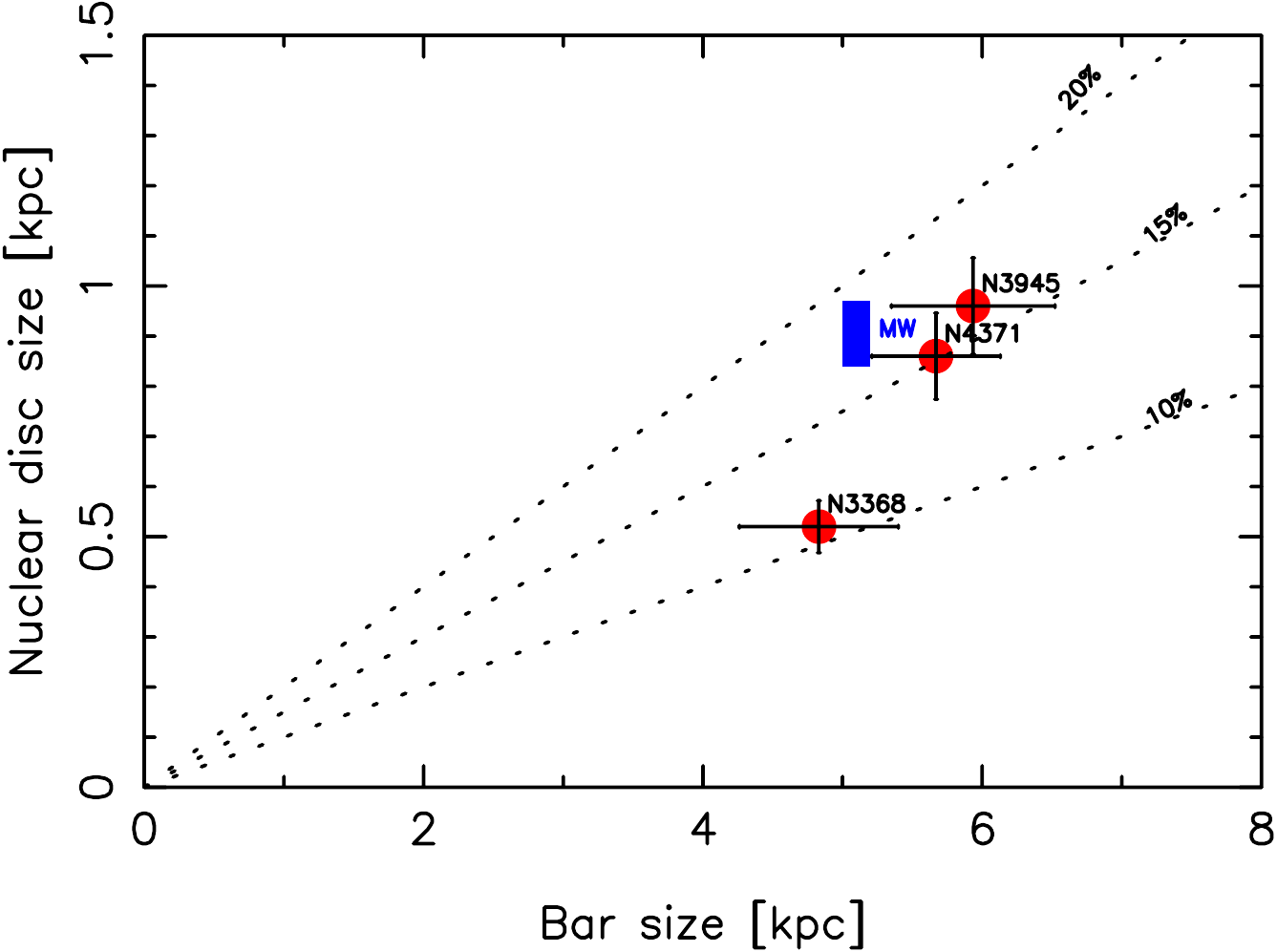}
\caption{Comparison of the size of the proposed nuclear disc and bar
  semi-major axes for the MW (blue rectangle, with sides indicating
  the respective uncertainties) and the sample of galaxies presented
  by \citet{cole+14} which all host perpendicular nuclear structures
  (red circles).  Dotted lines indicate constant fraction of bar size,
  as indicated.  The bar size for the MW is from \citet{wegg+15}.}
\label{fig:externalgals}
\end{figure}

In D15 we estimated that the nuclear disc needed to explain the
high-\vgsr\ peaks in the MW would have a semi-major axis of order
$1\kpc$.  Here we refine this estimate.  Using APOGEE Data Release 12,
D15 found tentative evidence of a high velocity peak at $l=8\degrees$;
in Data Release 13, \citet{zhou_shen+17} find no evidence of a
high-\vgsr\ peak in this field.  Therefore we can now assume that the
line of sight at $l=6\degrees$ is tangent to the nuclear disc/ring.
We further assume that the nuclear disc/ring has an ellipticity in the
range $0 \leq e \leq 0.2$.  For a bar angle of $27\degrees$ to the
line of sight \citep{wegg_gerhard13} and a nuclear structure
orthogonal to the bar, we obtain a size $0.84$ to $0.97$ \kpc,
assuming a distance to the Galactic Centre of $8\kpc$.  In
Fig. \ref{fig:externalgals} we compare this size to the nuclear
structures observed in three galaxies in which the bar is observed
almost perpendicular to the line of nodes, which is the optimal
orientation for detecting a nuclear disc/ring orthogonal to the bar.
The proposed nuclear disc/ring is similar in size, as a fraction of
its bar size, to these galaxies and is therefore not unreasonably
large.  Moreover one of the galaxies in this sample, NGC~4371, has a
known nuclear disc age of $\sim 11\Gyr$ \citep{gadotti+15}, further
demonstrating that such structures built from x2 orbits are stable
over long periods. \\

Therefore unlike the model of AS15, the high velocity peaks produced
by x2 orbits do not require the presence of preferentially young
stars.  Since APOGEE has not found young stars in the high velocity
peaks \citep{zasowski+16, zhou_shen+17}, this makes the x2 model a
very promising model for explaining these peaks in the MW.


\section{The Simulation}
\label{sec:simulation}

The simulation we use here is the same one we used in D15, which was
described more fully in \citet{cole+14}.  \citet{cole+14} found that a
nuclear disc formed in this simulation, which they showed is
qualitatively similar to the nuclear discs in three early-type
galaxies.  \citet{ness+14} also studied the same model to compare the
stellar age distribution in the bulge with that of the MW.
Lastly, \citet{debattista+17} analysed this model in some detail to
demonstrate that all the trends seen in the MW's bulge can be
understood as arising from internal evolution via the process of {\it
  kinematic fractionation}.  We therefore only provide a brief
description of the simulation here, and refer to those papers for
further details.

The simulation was evolved with the $N$-body$+$smooth particle
hydrodynamics code {\sc gasoline} \citep{gasoline}.  It starts with 5
million gas particles and 5 million dark matter particles.  This high
mass resolution allows us to use a softening of 100 \pc\ for the dark
matter particles and 50 \pc\ for the gas and stellar particles.  In
the simulation, gas from a hot corona in pressure equilibrium with a
dark matter halo cools and settles into a disc.  At high gas density
(greater than 100 amu $\mathrm{cm}^{-3}$) star formation is triggered;
thereafter feedback from star particles is provided via asymptotic
giant branch star winds and Types Ia and II supernovae.  We use the
blastwave prescription of \citet{stinson+06} to model the supernovae
feedback.  The gas particles initially all have a mass of $2.7 \times
10^4 \Msun$ and stars are born with $35\%$ of this mass.  Once the
mass of gas particles drops below $21\%$ of their starting value, they
are removed and their mass is distributed to the nearest neighbours.
By $10~\Gyr$ the simulation has formed $\sim 1.1\times 10^7$ star
particles with a total mass of $6.5 \times 10^{10} \Msun$.

\subsection{Model scaling and sampling}
\label{ssec:scaling}

We adopt the same scaling of the model as in D15 to facilitate
comparison with that work.  \citet{cole+14} showed that, after 6 Gyr,
the model forms a prominent nuclear disc encircled by an elliptical
star-forming ring.  By $10~\Gyr$ the nuclear disc has a semi-major
axis of $1.5~\kpc$ and is quite massive and unlikely to match any
nuclear disc in the MW.  Therefore we consider the model at
$7.5\Gyr$ (referred to as \tm{2}\ in D15) when a strong nuclear disc
is established.

The model is more compact and rotates more rapidly than the MW;
as in D15, we therefore rescale it in size and velocity.  The bar has
a size of $\sim 2.1\kpc$; assuming the MW's bar is 3.5 kpc long
\citep{gerhard02} D15 rescaled all coordinates by a factor of 1.67.
D15 rescaled the velocities using a least-squares match of the
line-of-sight velocity dispersion of the model to ARGOS survey data
\citep{ness+13b} for all stars within Galactocentric radius $R_{\rm
  GC} < 3.5$ \kpc\ at $b$ = $5\degrees$, $7.5\degrees$ and
$10\degrees$ across $|l| < 15\degrees$.  This gives a scaling factor
of 0.48 for velocities.  While position and velocity scalings lead to
the model becoming somewhat similar to the MW, it remains a not
very good match to the MW.

We place the observer at $y=-8$ \kpc, and orient the bar at
$27\degrees$ to the line of sight \citep{wegg_gerhard13}.  We adopt a
selection function for star particles in the model:
\begin{equation}
P(\rsun) = 
\begin{cases}
 w(A) & \text{for } 2 \kpc \leq \rsun \leq 10 \kpc, \\
 0 & \text{otherwise,}
\end{cases}
\end{equation}
where \rsun\ is distance from the Sun, and $w(A)$ is an age-dependant
weight.  In most cases we set $w(A) = 1$ for all ages, as in D15.  But
we also consider cases where we reduce the weight of just the younger
stars, setting $w(A) = 0.1$ or $w(A) = 0.2$ for stars younger than
$1\Gyr$, to compensate for the high star formation rate.  D15
presented an example of $w(A) = 0.2$ for stars younger than $1\Gyr$,
which stars in the nuclear disk are.  As in D15, when considering
distributions of kinematic observables we use an opening angle of
$0.5\degrees$ which matches the smallest size of the APOGEE fields.

\subsection{Limitations of the model}

The model is useful for interpreting and predicting trends in future
data, but it should not be construed as a detailed model of the MW,
even after it is rescaled.  Its primary advantage is that it is one of
the first simulations with gas and star formation where all the stars
are formed from gas and where a nuclear disc forms.  On the other
hand, it has a number of limitations which should give pause to any
efforts to test the model on a detailed quantitative basis.

Foremost of the limitations is that the resolution used for the gas is
still too large to properly resolve the gas ring size.
\citet{sormani+15} show, using two-dimensional grid calculations with
a fixed bar potential, that the size of the gas ring that forms is
dependent on the grid cell size, varying by a factor of $\sim 2$ when
this cell size is changed from $40\pc$ to $10\pc$ \citep[see
  also][]{lizhi+15}.  They interpret this variation as resulting from
the need to resolve the cusped x1 orbit, at which point the gas shocks
and falls inwards onto x2 orbits \citep{binney+91, sormani+15}.  This
is primarily a hydrodynamical problem, not one of force resolution.
The finite number of particles needed for the SPH kernel results in
the gas shocking and transitioning from x1 to x2 orbits at too large a
radius, as described by \citet{sormani+15}.  Thus there is every
reason to believe that the ring in our simulation, and therefore the
nuclear disc that forms from it, is too large, as was already noted by
\citet{cole+14}.  While the gas disc is too large, this does not mean
that the extent of the x2 orbits that support the stellar nuclear disc
is too large.  The extent of the x2 orbits is set by the gravitational
potential which is well resolved on the scale of the stellar disc
(corresponding to $20$ gravitational softening lengths).  It is only
how far out on the x2 orbits that the gas settles onto that is at
question.  \citet{lizhi+15} show that the extent of x2 orbits is
considerably larger than the region where gas settles.  Our simulation
populates x2 orbits because of still too low mass resolution, but in
the MW x2 orbits may have been populated in other ways.  For
instance it is possible that external perturbations may give rise to
gas settling on such orbits and forming stars.

One of the consequences of the large nuclear disc is that we need to
scale the model to an old bar size for the MW, 3.5~\kpc.  Scaling to
\citet{wegg+15}'s bar size ($5 \kpc$) results in too large a nuclear
disc.  This does not imply that the nuclear disc in the MW favours the
smaller bar size; indeed our scaling results in a nuclear disc that is
still too large to match the MW, with \vgsr\ peaks at slightly larger
longitudes than in the MW (D15).

Another issue that arises is that the star formation rate (SFR) in the
simulation's nuclear disc is very high.  \citet{cole+14} estimate a
SFR of $\sim 1.5 \Msun\yr^{-1}$ within $1\kpc$.  The MW's Central
Molecular Zone is forming stars at a rate of $\sim 0.14 \Msun\yr^{-1}$
\citep{wardle_yusef-zadeh08}.  As a result the model rapidly builds up
to a mass much higher than D15 estimated for a MW nuclear disc.
Because of this difference, the stellar nuclear disc is much more
prominent in the model's LOSVDs than the high-\vgsr\ peaks in the
APOGEE data.  D15 present an example of artificially reducing the
contribution of young stars by a factor of 5.  The resulting LOSVD
high-\vgsr\ peak is more realistic relative to the main,
low-\vgsr\ peak.  Below we therefore also show the results of reducing
these weights, by setting $w(A) = 0.1-0.2$.\footnote{However this
  factor of 10 difference is only with the {\it current} star
  formation rate of the MW.  We argue that a MW nuclear disc/ring may
  be quite old, and the MW's star formation rate a earlier times was
  probably higher.  A factor of 10 is therefore probably an
  over-estimate of the amount by which young, nuclear disc stars need
  to be downweighted for a more realistic comparison with the MW.}

Thus the model is not a good match to the MW.  {\it Therefore our
  predictions from the model are qualitative trends rather than exact
  kinematic values or locations of features.}


\section{Signatures of a Nuclear Disc}
\label{sec:signatures}

\subsection{Density distribution}

\citet{cole+14} showed that the nuclear disc in the model is
elliptical and orientated perpendicular to the main bar, because it is
supported by x2 orbits (Earp et al. {\it in progress}).  Since the
near side of the bar in the MW is at positive longitudes, the near
side of the nuclear disc is at negative longitudes.  Thus it should be
detectable at larger longitudes at $l < 0\degrees$ compared with $l >
0\degrees$.  However because the nuclear disc is rounder than the bar,
with ellipticity $\sim 0.2$ in the simulation, and because the nuclear
disc is almost side-on, the density difference between positive and
negative longitudes is small and would be hard to detect.

\subsection{Proper motions}

\begin{figure*}
\centerline{
    \includegraphics[angle=0.,width=0.45\hsize]{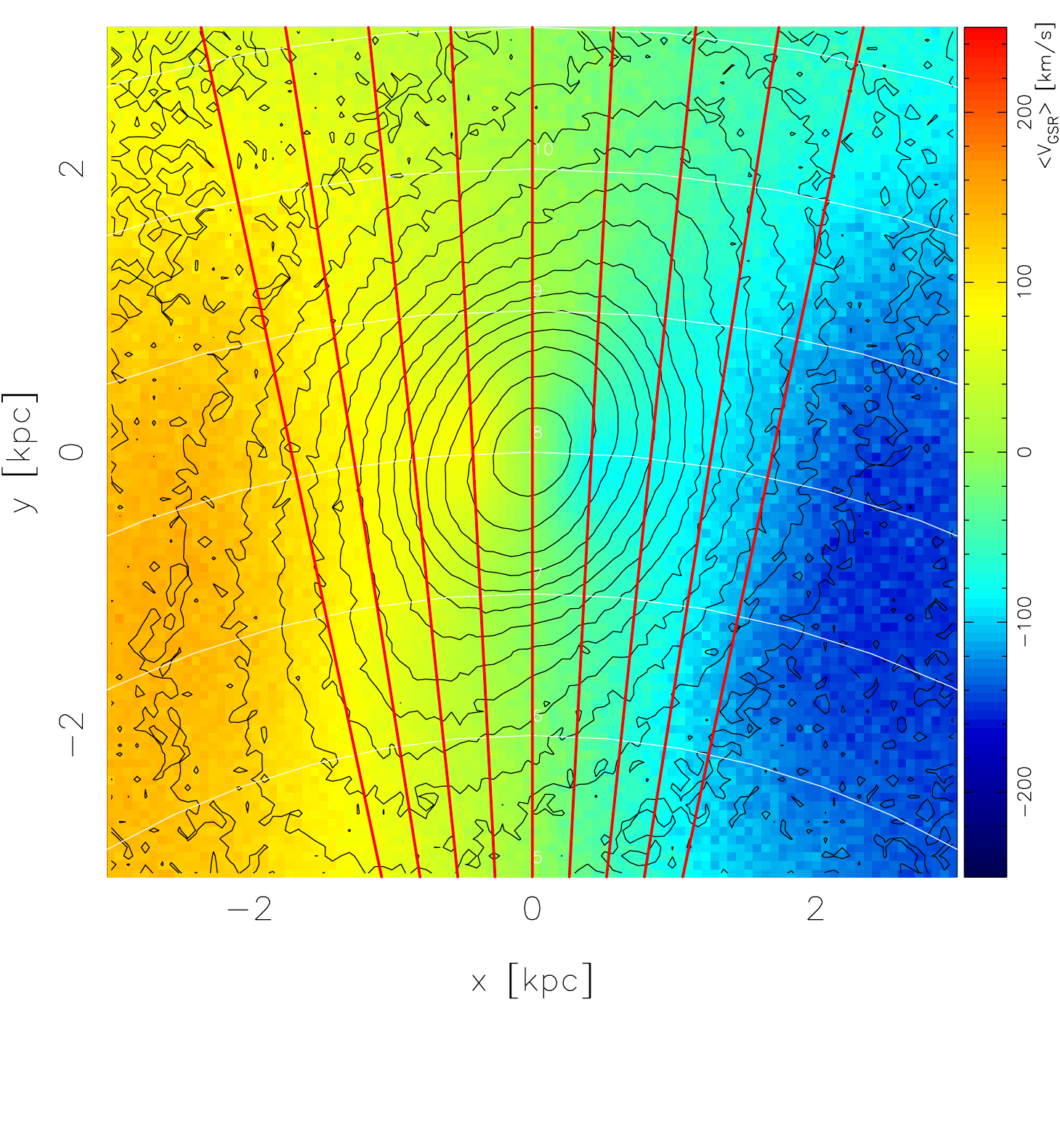}
    \includegraphics[angle=0.,width=0.45\hsize]{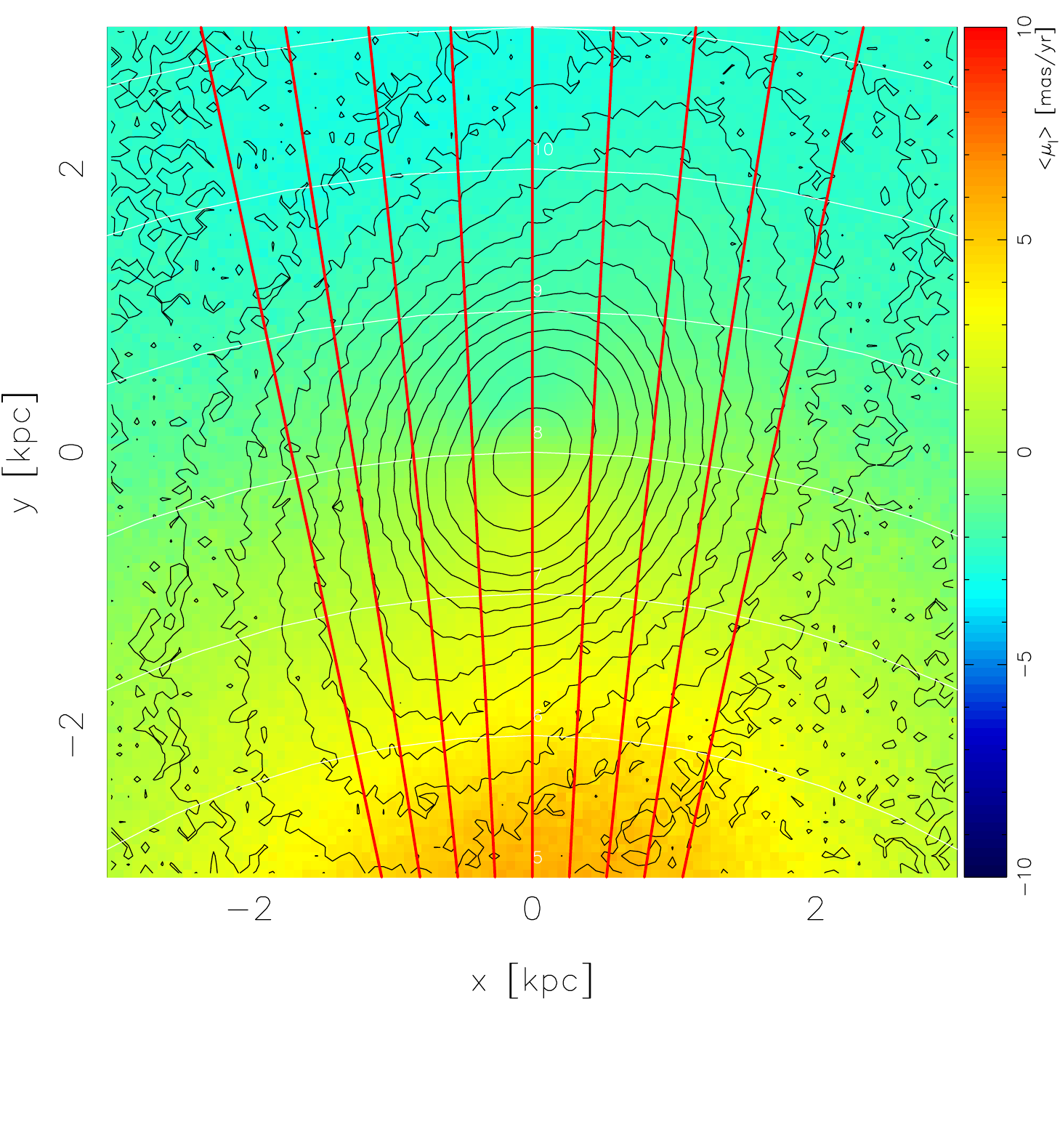}
}
\centerline{
    \includegraphics[angle=0.,width=0.45\hsize]{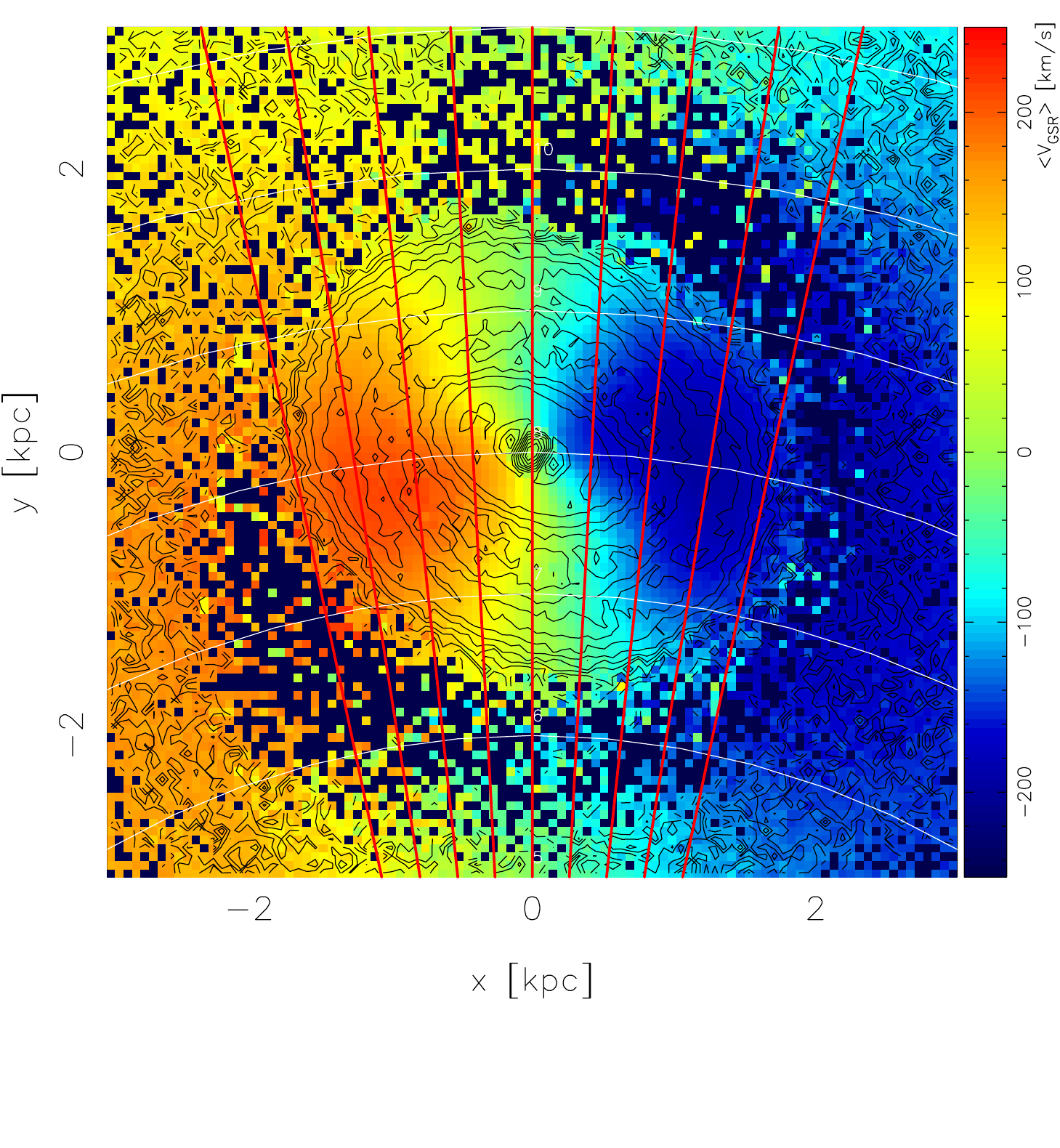}
    \includegraphics[angle=0.,width=0.45\hsize]{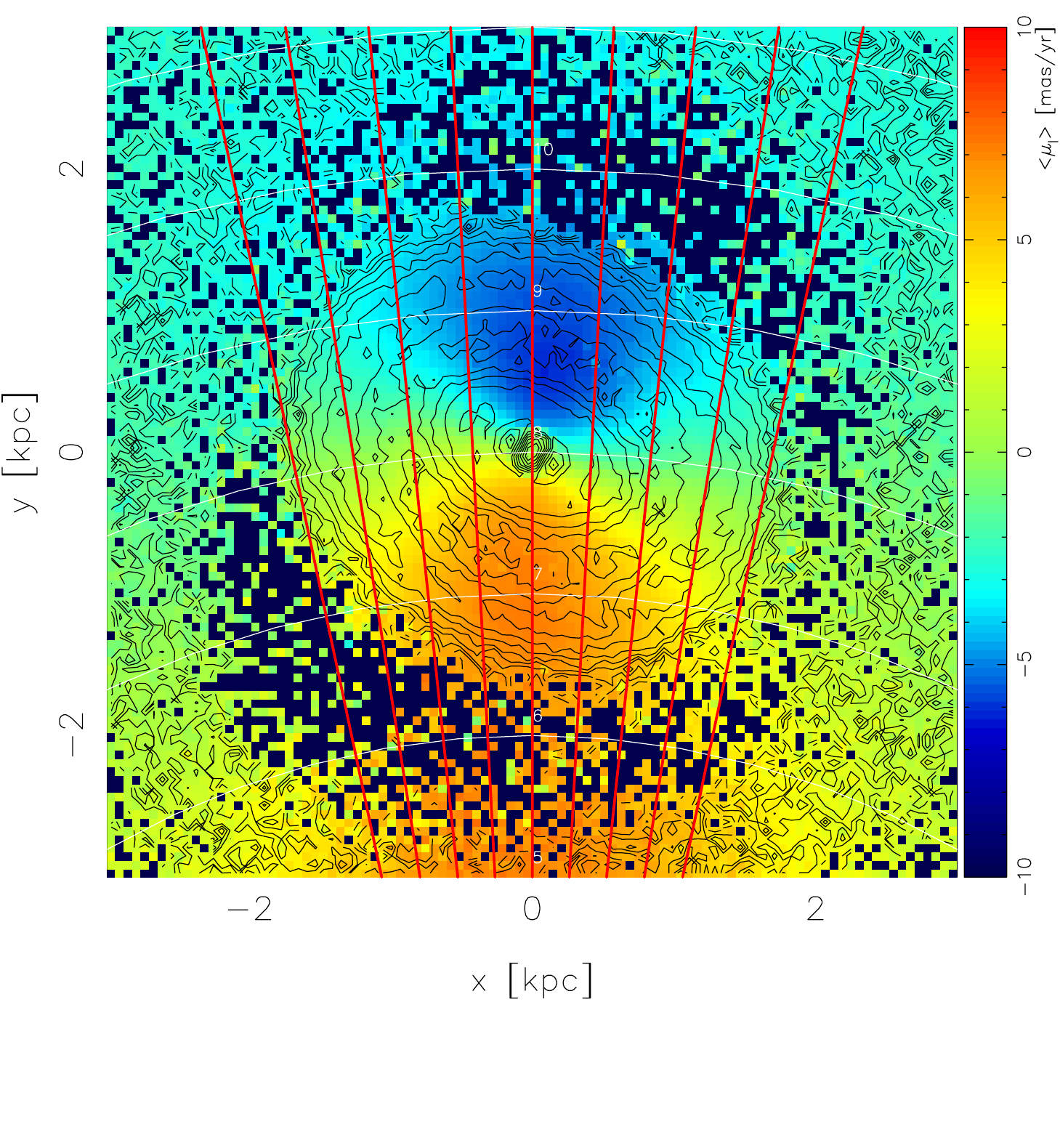}
}
\caption{Kinematics of the model for predominantly bar (age $> 2\Gyr$)
  stars (top row) versus predominantly nuclear disc (age $< 0.5 \Gyr$)
  stars (bottom row) as seen from the Sun ($x,y = 0,-8$\kpc).  Left
  panels show \avg{\vgsr}, while right panels show \avg{\pmo{l}}.
  White circles indicate constant distance from the Sun, \rsun, while
  red lines show $l = 0\degrees$, $\pm 3\degrees$, $\pm 6\degrees$,
  $\pm 9\degrees$, and $\pm 12\degrees$.}
\label{fig:kinemapsage}
\end{figure*}

\begin{figure*}
\includegraphics[angle=0.,width=0.9\hsize]{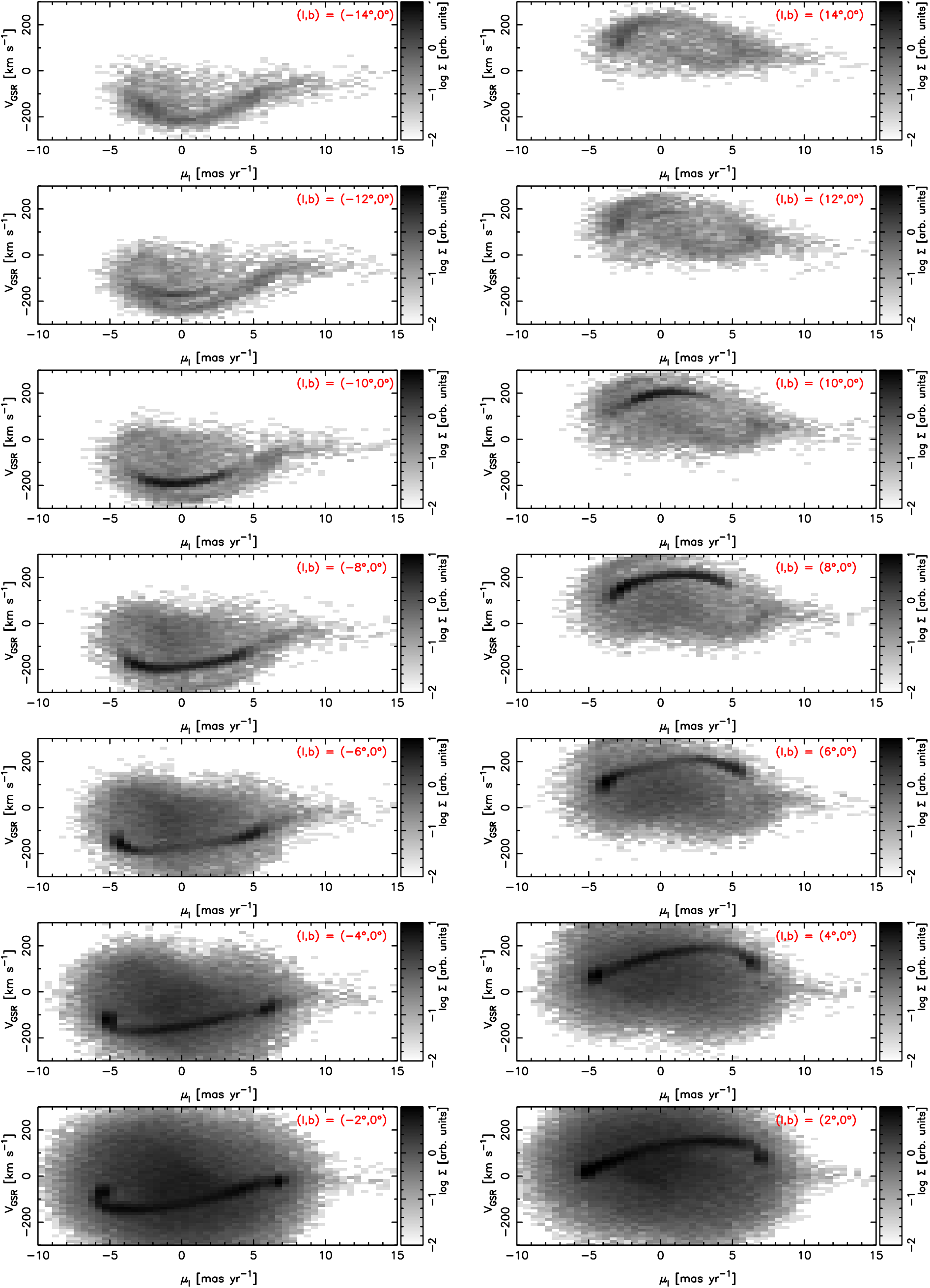}
\caption{\pmo{l}\ versus \vgsr\ of the model for different lines of
  sight in the mid-plane.}
\label{fig:2dkinematics750a}
\end{figure*}

We turn therefore to the kinematics to search for evidence of an
elliptical nuclear disc.  \citet{cole+14} showed that the kinematics
of stars in the bar and in the nuclear disc, which they separated by
means of an age cut, are different.  Stars in the bar stream along the
bar.  As seen from the centre of the galaxy, these stars have negative
(infalling) radial motions on the leading edge of the bar; nuclear
disc stars instead have negative radial motions on the trailing side
of the bar.  This pattern is produced because the nuclear x2 disc is
elongated perpendicular to the bar.

We consider two components of stellar motions as seen from the Sun:
\vgsr, the Galactocentric radial velocity and \pmo{l}, the proper
motion in the $l$ direction in the Galactic rest frame.  The nuclear
disc manifests as an over-density of stars with low \pmo{b}, the
proper motion in the $b$ direction, as is expected for a thin disc.
However the parallax differences between the positive and negative
longitude sides of a nuclear disc would be too small to be detectable,
rendering vertical proper motions of relatively limited use for
understanding the structure of a nuclear disc.  We therefore consider
the signature of a nuclear disc only in the space spanned by
\vgsr\ and \pmo{l}.  This space also allows easy interpretation of the
LOSVDs that we present in the next section.
Fig. \ref{fig:kinemapsage} presents maps of \avg{\vgsr}\ and
\avg{\pmo{l}}, for predominantly bar stars (top row) and predominantly
nuclear disc stars (bottom row), via an age cut.  The streaming
motions in the nuclear disc are larger overall.  In the nuclear disc
\avg{\pmo{l}}\ is very large and distinct from that of the main
bar. Likewise \avg{\vgsr}\ is large to smaller $|l|$ than in the bar
stars.  These different kinematics as seen from the Sun allow a
nuclear disc to be recognised.

Fig. \ref{fig:2dkinematics750a} shows the distribution of stars in the
model's \pmo{l}-\vgsr\ plane at $b=0\degrees$ across $-14\degrees \leq
l \leq 14\degrees$.  The nuclear disc is evident as a narrow,
continuous distribution at $|l| \la 10\degrees$ surrounded by a sea of
bar and disc particles which produce the large spreads in \vgsr\ and
\pmo{l}.  As shown by D15, the high-\vgsr\ peaks are absent at $|b| =
2\degrees$ and above, and we also find no sign of the nuclear disc in
the \pmo{l}-\vgsr\ plane at this latitude.  Any high velocity peaks
observed in this region \citep[e.g.][]{nidever+12, zhou_shen+17} must
therefore have an origin other than a nuclear disc.

\begin{figure*}
\centerline{
    \includegraphics[angle=0.,width=0.45\hsize]{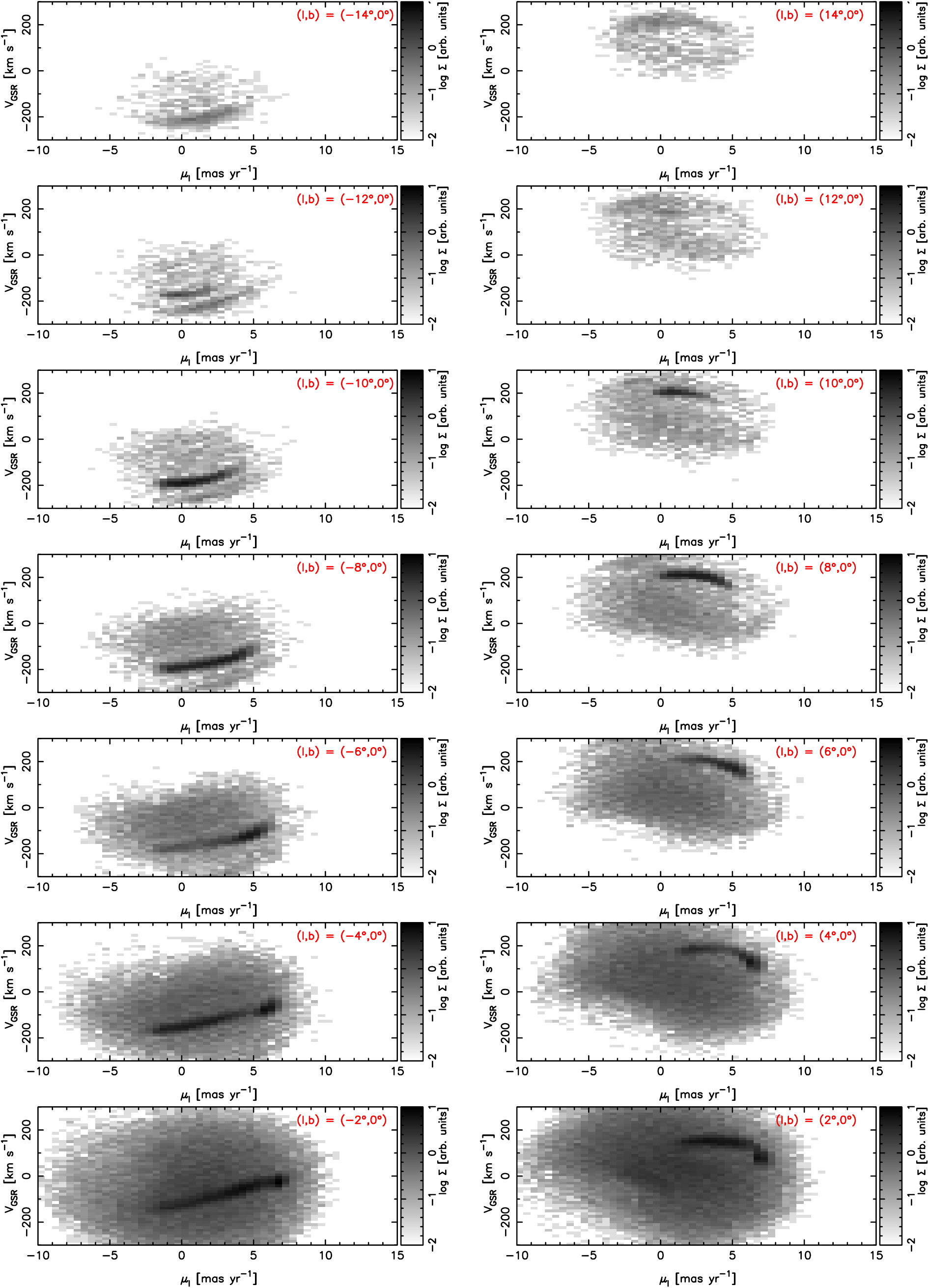}
    \includegraphics[angle=0.,width=0.45\hsize]{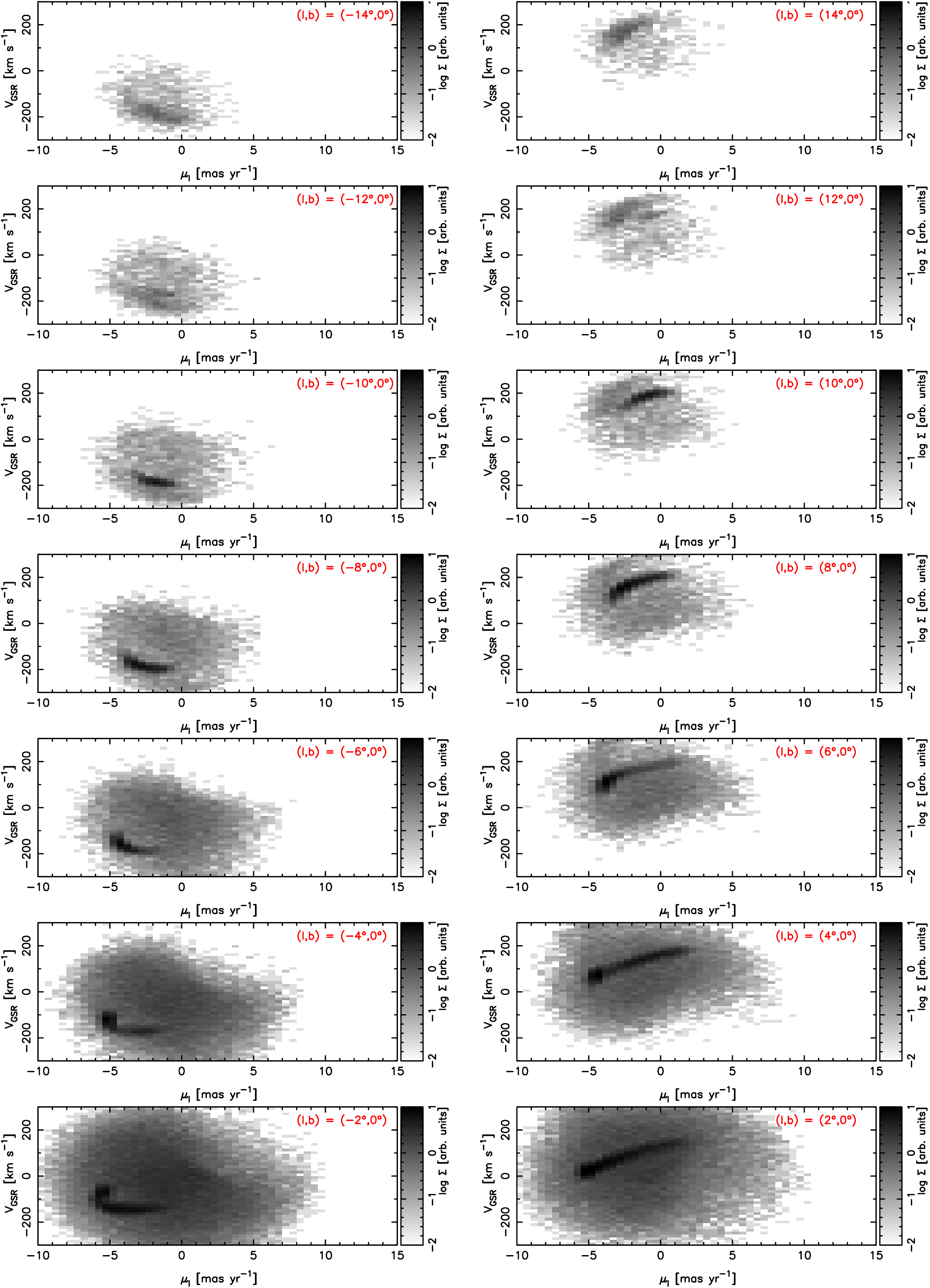}
}
\caption{\pmo{l}\ versus \vgsr\ of the model for different lines of
  sight in the mid-plane. Left two columns: near side ($6.5 \leq
  \rsun/\kpc \leq 8$) stars.  Right two columns: far side ($8 \leq
  \rsun/\kpc \leq 9.5$) stars.}
\label{fig:2dkinematics750byrad}
\end{figure*}

Fig. \ref{fig:2dkinematics750byrad} deconstructs the trace of the
model's nuclear disc in the \pmo{l}-\vgsr\ plane as a function of
\rsun.  The near side ($\rsun < 8\kpc$) contributes the $\pmo{l} > 0$
part of the track, with the distant side providing the negative
$\pmo{l}$ part.  The near side ($\rsun < 8\kpc$) of the nuclear disc
is at $\pmo{l} > 0$, which results in the largest absolute proper
motions being positive.

A nuclear disc can be seen to be non-axisymmetric in two ways in the
\pmo{l}-\vgsr\ plane.  First, its track is not symmetric about
$\vgsr=0$ when comparing positive and negative $l$, indicating that
the nuclear disc is neither circular nor perpendicular to the line of
sight.  The nuclear disc also can be seen to not be circular from just
a single $l$.  The nuclear disc's track reaches $\pmo{l} = 0$ at the
radius where stars are moving radially towards or away from the Sun.
If the nuclear disc were circular, at $\pmo{l} = 0$ the line of sight
would be at the smallest Galactocentric radius within the nuclear
disc; the slope of the track would therefore be flat.  The clear slope
at $\pmo{l} = 0$ is therefore a sign that a nuclear disc is not
axisymmetric.  The slope ${\mathrm d} \vgsr/{\mathrm d}\pmo{l}$ is
positive because of the orientation of the bar, which places the far
side of the nuclear disc at positive $l$; if the bar angle to the line
of sight had been negative, then the slope would be negative.

\begin{figure*}
\includegraphics[angle=0.,width=0.9\hsize]{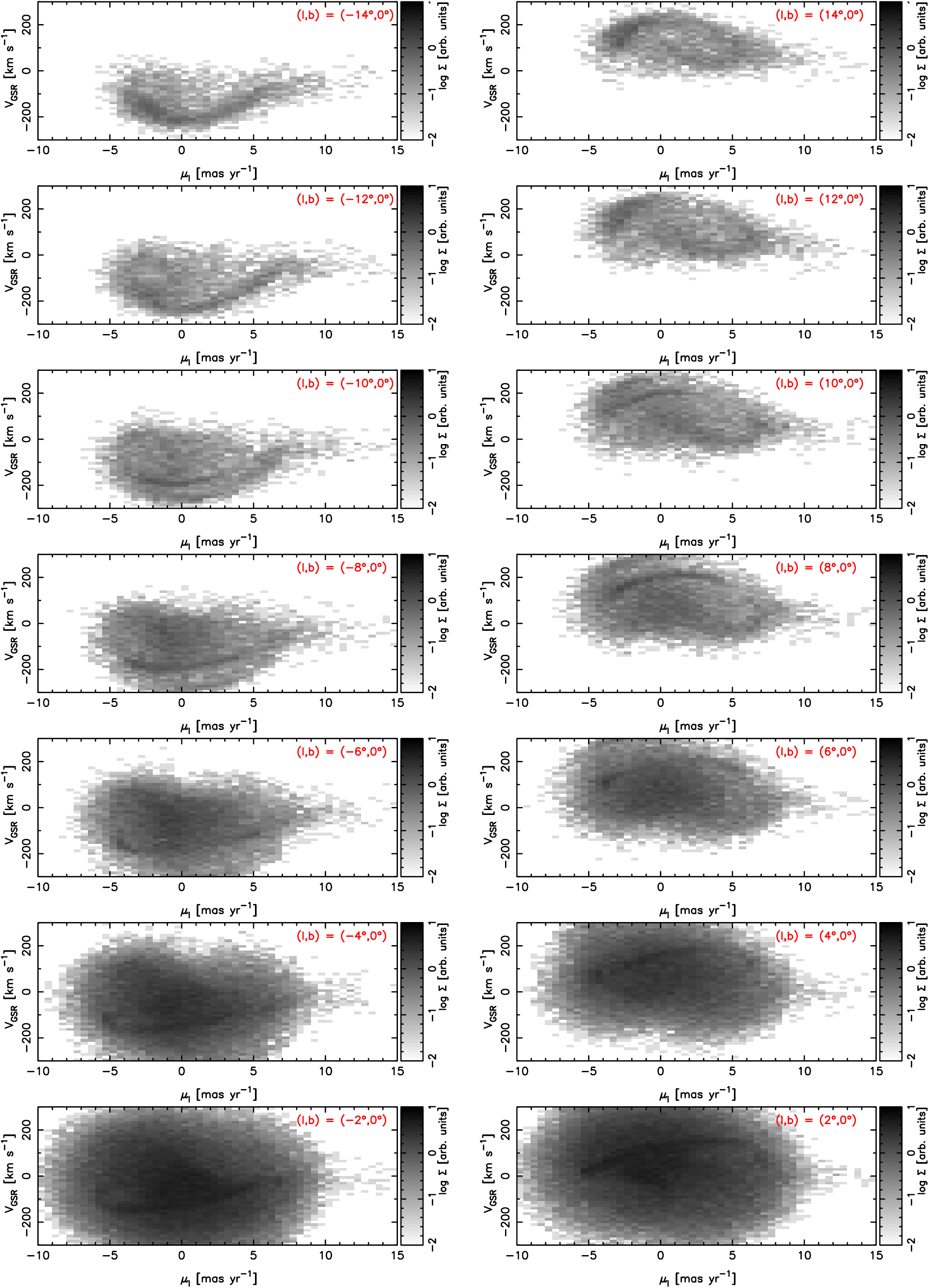}
\caption{\pmo{l}\ versus \vgsr\ of the model for different lines of
  sight in the mid-plane with $w(A) =0.1$ for stars younger than
  $1\Gyr$.}
\label{fig:2dkinematics750downweighted}
\end{figure*}

The nuclear disc in our model is relatively massive, which means that
its trace in the \pmo{l}-\vgsr\ plane is clearer than would be the
case in the MW.  In order to test the significance of the nuclear disc
trace at a more realistic mass level, in Fig.
\ref{fig:2dkinematics750downweighted} we adopt $w(A) = 0.1$ for stars
younger than $1\Gyr$.  The nuclear disc trace is harder to distinguish
in large parts of the space, but remains evident at $l=\pm 8\degrees$
and $\pm 10\degrees$.

\subsection{Line-of-sight velocity distributions}

\begin{figure}
\centerline{
\includegraphics[angle=0.,width=1.\hsize]{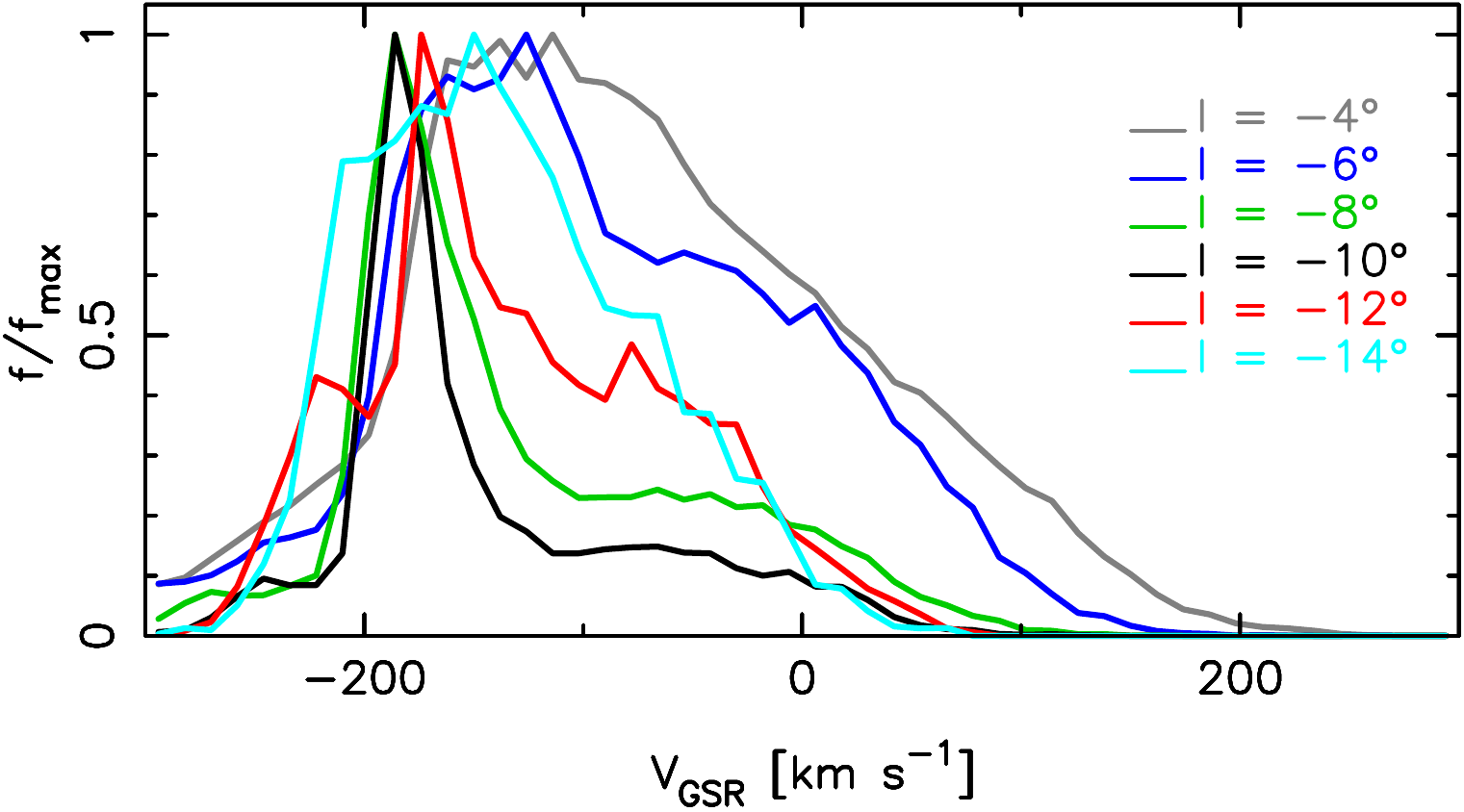}\
}
\centerline{
\includegraphics[angle=0.,width=1.\hsize]{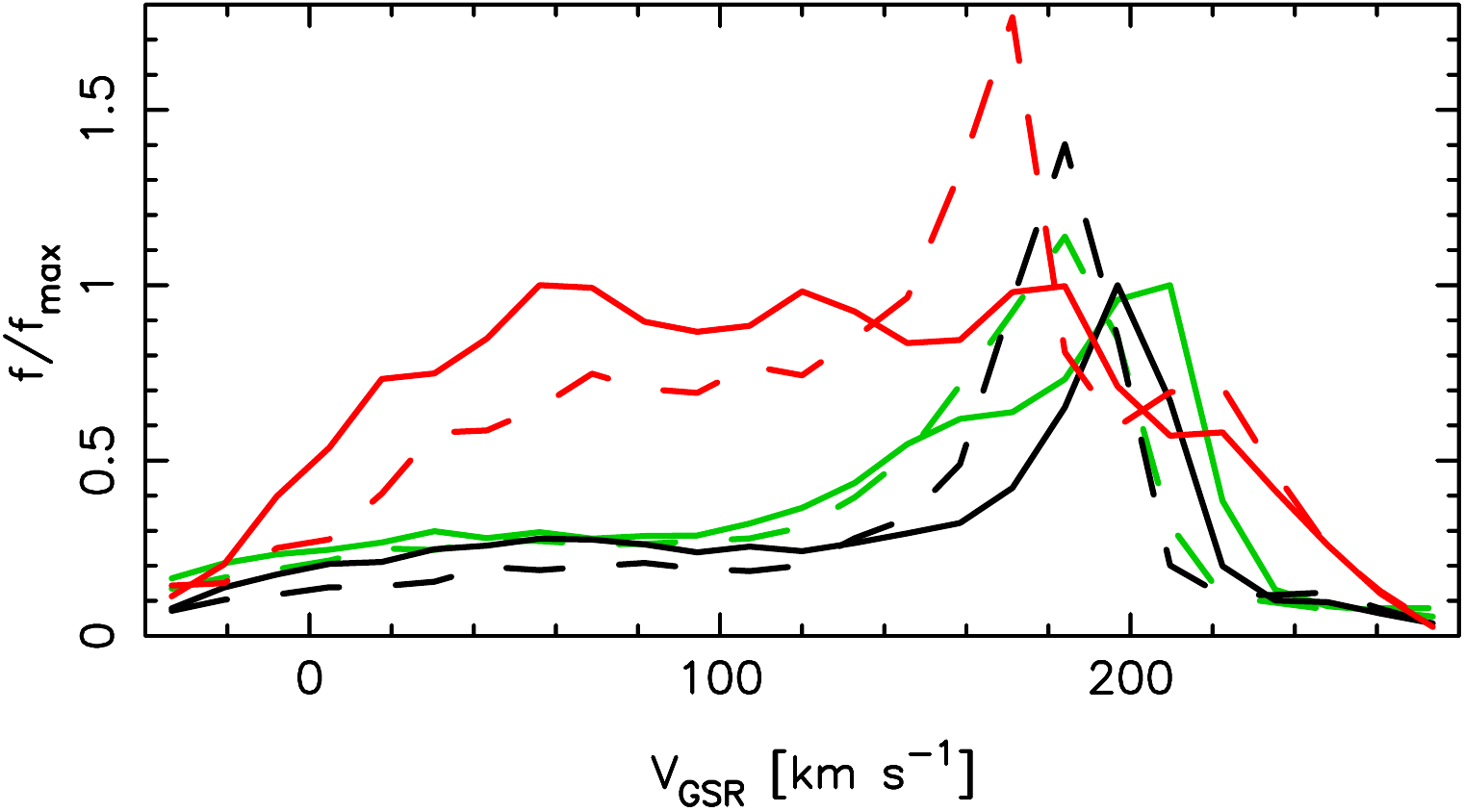}\
}
\centerline{
\includegraphics[angle=0.,width=1.\hsize]{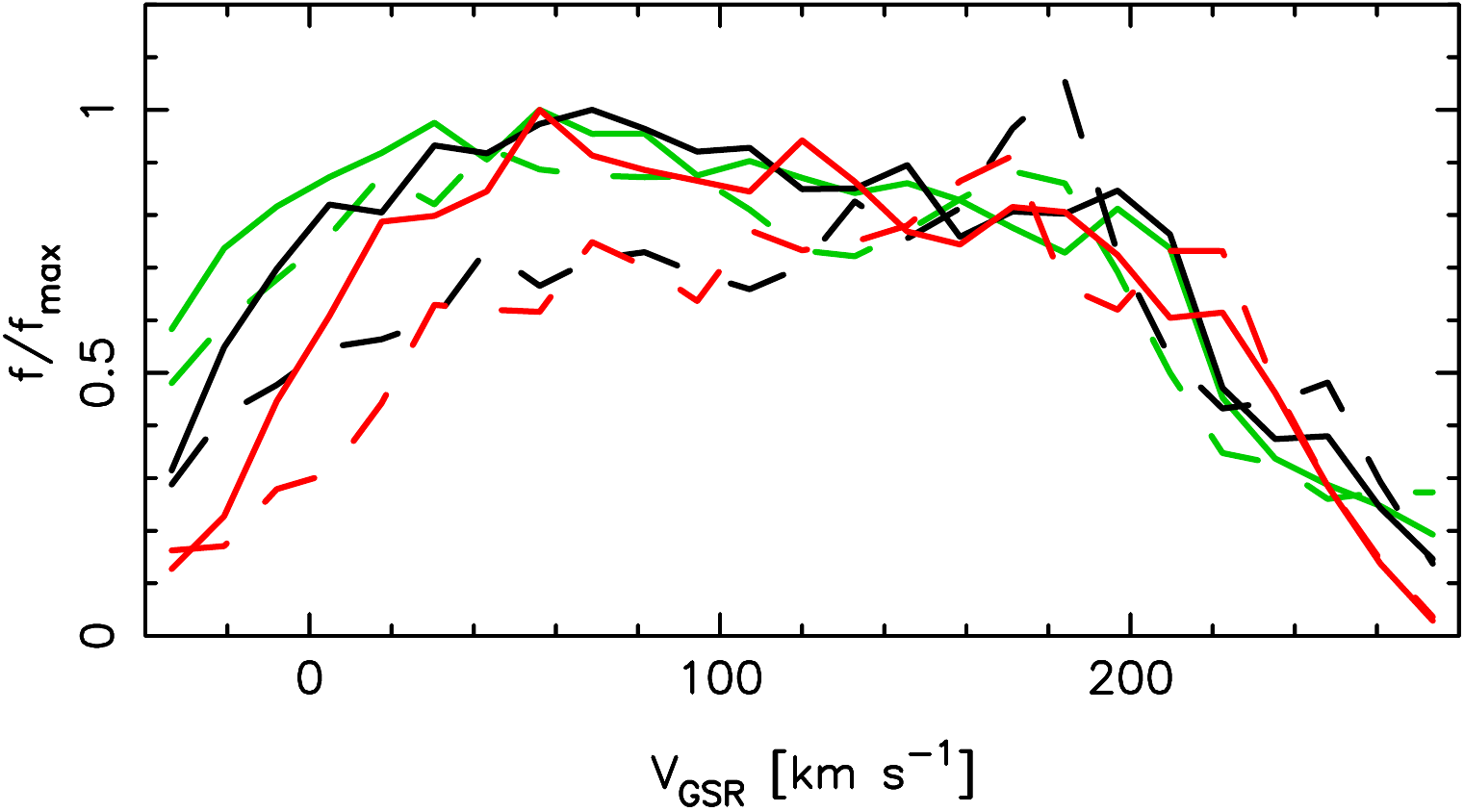}\
}
\centerline{
\includegraphics[angle=0.,width=1.\hsize]{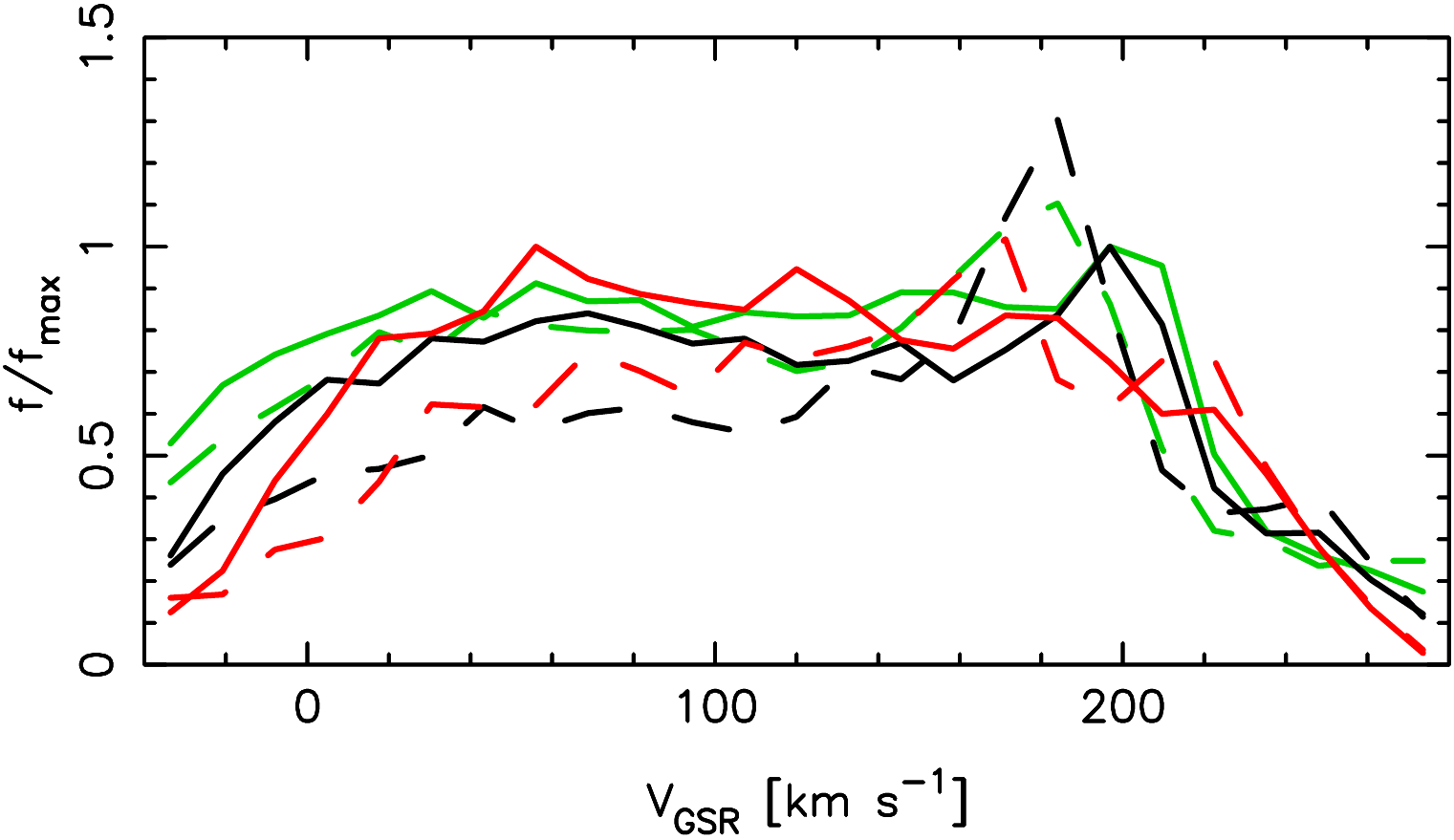}\
}
\caption{Top: Mid-plane LOSVDs for the model at negative longitudes,
  as indicated.  Each LOSVD has been normalised to unit peak.  Second
  row: Comparison between the model's mid-plane LOSVDs at negative and
  positive longitudes.  Solid lines show $l > 0\degrees$ while dashed
  lines show $l < 0\degrees$ (with the sign of \vgsr\ reversed).  At
  each longitude, the LOSVDs are normalised to the peak at positive
  $l$.  Colours indicate $|l|$ as in the top panel.  Third and fourth
  rows: Identical to second row with $w(A) = 0.1$ and $0.2$,
  respectively.}
\label{fig:losvds}
\end{figure}

While proper motions in even a single line of sight crossing a nuclear
disc provide considerable diagnostic information, they are still
challenging to measure observationally, so we now turn to the LOSVDs.
LOSVDs are projections onto the \vgsr\ axis of
Fig. \ref{fig:2dkinematics750a}.  These projections often obscure the
nuclear disc except in a few select directions.  D15 compared the
model to APOGEE data at $l > 0\degrees$ and showed some of the
signatures by which a nuclear disc can be recognised.  A nuclear disc
is evident in LOSVDs at positive longitudes as a high-\vgsr\ peak,
which is cooler (narrower) than the dominant low-\vgsr\ peak, and is
absent off the mid-plane.  Moreover, the distribution around the
high-\vgsr\ peak is skewed towards smaller \vgsr.  D15 found that the
first three properties are matched by the APOGEE LOSVDs at
$l=6\degrees$ to $8\degrees$.  However, the signal-to-noise ratio
(number of stars observed) was not sufficiently high to make
definitive statements about the skewness.

Fig. \ref{fig:losvds} presents the LOSVDs of the model at negative
longitudes.  The top panel of Fig. \ref{fig:losvds} shows the LOSVDs
at the opposite longitudes to those in D15 (their Fig. 1).  At
negative longitudes, we find the same four basic predictions with some
differences characteristic of an elliptical nuclear disc.  The first
obvious difference produced by the ellipticity of a nuclear disc is
that the high-\vgsr\ peak extends to larger $|l|$ on the negative side
compared with the positive side, because the near side of the nuclear
disc is at $l < 0\degrees$.

Fig. \ref{fig:2dkinematics750a} showed that a nuclear disc on the near
side of the bulge produces a high-\vgsr\ peak at $l>0\degrees$.
Correspondingly, the nuclear disc on the far side of the bulge
produces the high-\vgsr\ peaks at $l < 0\degrees$.  In general the
peak $|\vgsr|$ is larger at $l>0\degrees$ than at $l<0\degrees$,
because the line of sight crosses the nuclear disc closer to its minor
axis, where the velocity is larger.  The top panel of
Fig. \ref{fig:losvds} shows that the high-\vgsr\ peak in our model is
at $|\vgsr| \simeq 180\kms$ at $l<0\degrees$; in contrast at $l >
0\degrees$ the peak is at $\vgsr \simeq 200 \kms$.  Conversely,
because the lines of sight on the $l < 0\degrees$ side intersect the
nuclear disc at a smaller range of Galactocentric distances than at $l
> 0\degrees$, the $l < 0\degrees$ peak has a narrower range of
velocities, \ie\ it appears cooler, than at $l > 0\degrees$.

The second panel of Fig. \ref{fig:losvds} compares the LOSVDs at
negative and positive longitudes directly, transforming \vgsr\ to
$-\vgsr$ for the negative $l$ bins.  Aside from the high-\vgsr\ peak
occurring at smaller $|\vgsr|$ and being cooler, at negative
longitudes, two additional geometric effects are evident.  In the
low-\vgsr\ peak, the LOSVD is higher at $l>0\degrees$, which occurs
because the path length through the bar there is longer and intersects
the bar at smaller Galactic radii \citep{blitz_spergel91}.  For the
same reason, the high-\vgsr\ peak is more pronounced on the
$l<0\degrees$ side.  Therefore a high-\vgsr\ peak is easier to detect
at negative longitudes.  This asymmetry is a key signature of an
elliptical nuclear disc.

The third and fourth row of Fig. \ref{fig:losvds} are identical to the
second row but set $w(A) = 0.1$ and $0.2$, respectively, for stars
younger than $1\Gyr$.  Even in the case where $w(A) = 0.1$, which is
probably an over-estimate of the correction required, high
\vgsr\ peaks are evident, although the peak in the $l =-12\degrees$
bin is ambiguous.

\begin{figure}
\centerline{
\includegraphics[angle=0.,width=\hsize]{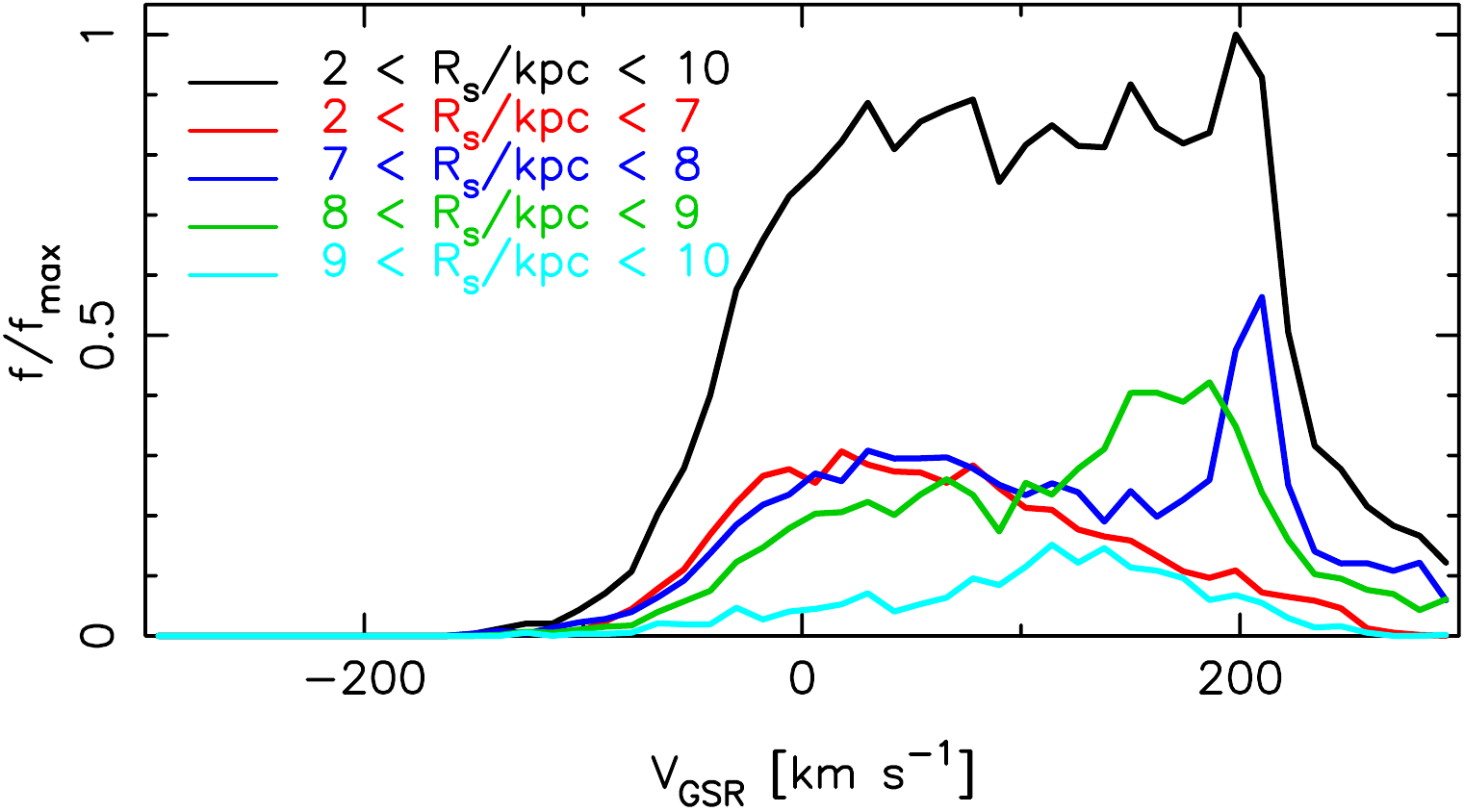}\
}
\centerline{
\includegraphics[angle=0.,width=\hsize]{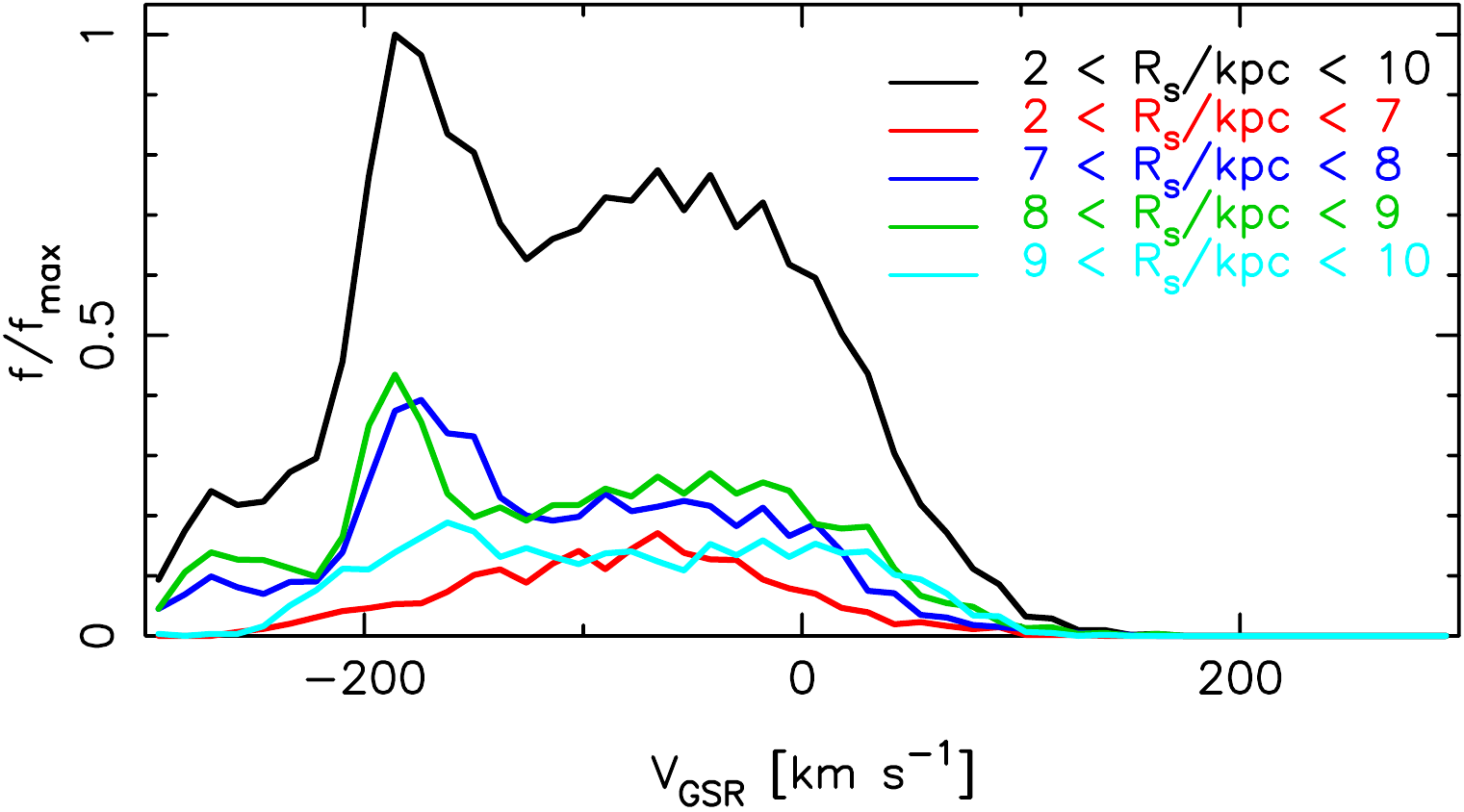}\
}
\caption{Top: decomposition by distance from the Sun of the model's
  LOSVDs in the fields at $l = +8\degrees$ (top) and $l = -8\degrees$
  (bottom) , with $w(A)$ set to 0.2 for stars younger than $1\Gyr$ for
  the sake of clarity.  In both panels the LOSVDs are normalized to
  the peak of the LOSVD for the full distance ($2 \leq \rsun/\kpc \leq
  10$) range in that longitude.}
\label{fig:distdecomp}
\end{figure}

Fig. \ref{fig:distdecomp} shows the contribution of stars at different
distances from the observer, \rsun, on the LOSVDs at $l =
\pm10\degrees$.  The stars in the high-\vgsr\ peak are all at $7 \leq
\rsun/\kpc \leq 9$.

\subsection{The signature of a ring}

\begin{figure*}
\includegraphics[angle=0.,width=0.9\hsize]{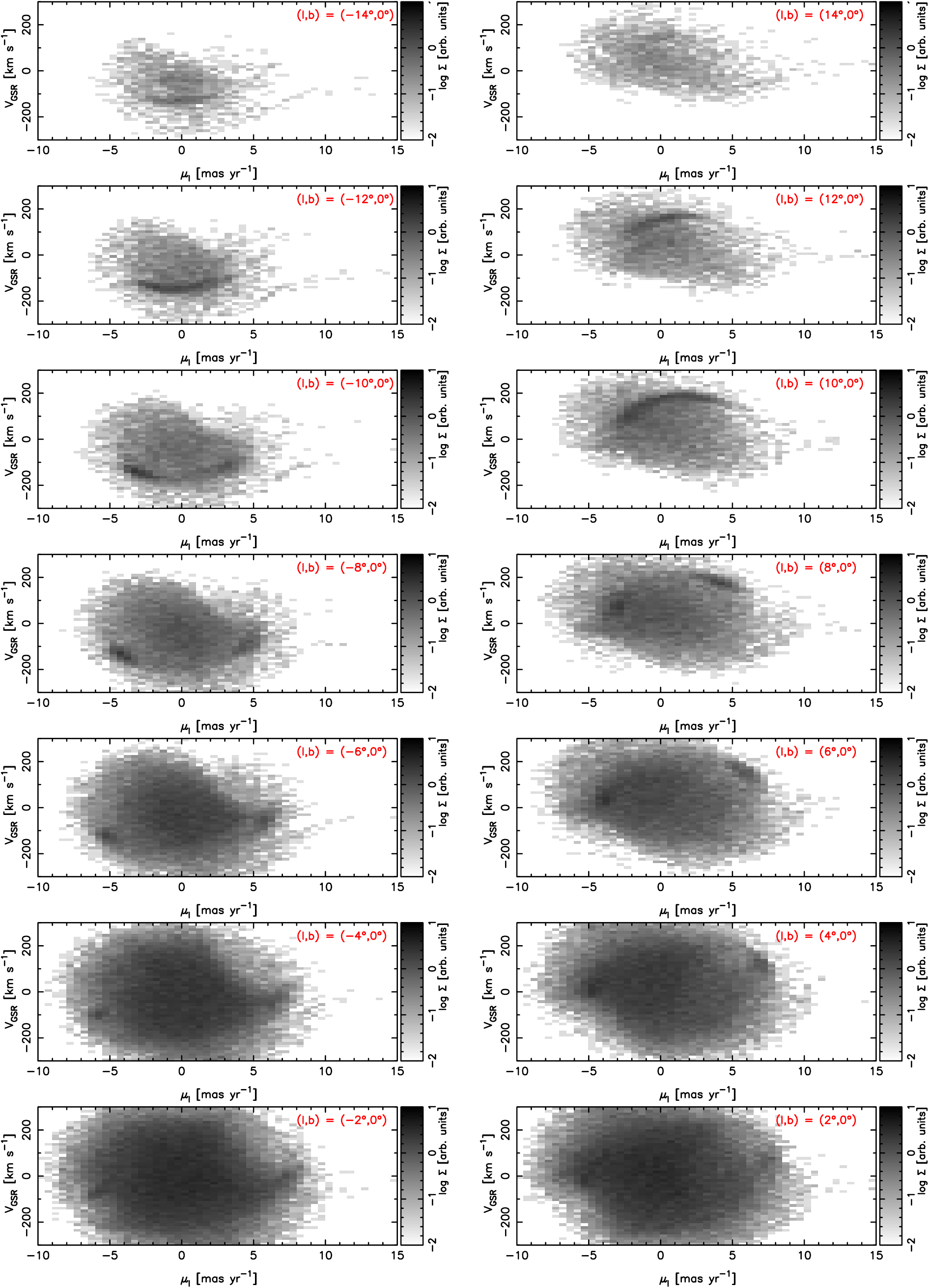}
\caption{\pmo{l}\ versus \vgsr\ for different lines of sight in the
  mid-plane for the simulation evolved from $7.5\Gyr$ to $13.5\Gyr$
  with gas cooling and star formation turned off.}
\label{fig:2dkinematicsNoSF}
\end{figure*}

While D15 interpreted the high-\vgsr\ peaks found by AGPOGEE as due to
a disc, a nuclear ring supported by x2 orbits is an equally viable
interpretation.  The fact that the nuclear loci in the
\pmo{l}-\vgsr\ plane are continuous shows that the nuclear structure
is a disc.  While the structure is continuous, a ring surrounds the
nuclear disc in our simulation and manifests as the peaks at the ends
of the nuclear disc track in the \pmo{l}-\vgsr\ plane.  This also
shows what the signature of a nuclear {\it ring} would be: two
disconnected overdensities in the \pmo{l}-\vgsr\ plane at most
longitudes, merging when the ring is seen tangentially.
Fig. \ref{fig:2dkinematicsNoSF} shows an example of a nuclear ring,
taken from the simulation evolved for $6\Gyr$ without star formation,
discussed in Section \ref{sec:ages}; the substantial evolution during
this time leads to the nuclear disc transforming into a ring.  This is
clear in the \pmo{l}-\vgsr\ plane as the disconnected pair of density
peaks seen at $l = \pm 6\degrees$ and at $l = 8\degrees$.  At $l = \pm
10\degrees$ and $l = \pm 12\degrees$ the nuclear ring is seen close to
tangentially and the trace in the \pmo{l}-\vgsr\ plane is continuous.


\section{Discussion} 
\label{sec:discussion}

We have presented further kinematic signatures of a nuclear disc at
the centre of the MW, particularly at negative longitudes, which will
be surveyed by APOGEE-2 and MOONS.  The principal signatures we
predicted are as follows:

\begin{itemize}

\item We showed that the LOSVDs contain second, high-\vgsr, peaks at
  $l < 0\degrees$ just as they do at $l > 0\degrees$.  The
  high-\vgsr\ peaks are cooler than the low-\vgsr\ peaks and skewed to
  smaller $|\vgsr|$, just as in the $l > 0\degrees$ case.

\item The LOSVDs at positive longitudes peak at larger $|\vgsr|$ but
  the peaks are visible to smaller $|l|$, than at negative longitude.
  These two properties are a result of the elliptical nature of a
  nuclear disc and the fact that it would be perpendicular to the bar.

\item Moreover the $l<0\degrees$ high-\vgsr\ peaks are higher than the
  $l>0\degrees$ ones.  Testing this property is only possible if the
  selection function is very well understood.

\item The proper motion in the latitude direction, \pmo{b}, shows only
  a narrow distribution, consistent with a population that is thin,
  which can be inferred already from the absence of high-\vgsr\ peaks
  off the mid-plane.  Any high-\vgsr\ peaks at $|b| \sim 2\degrees$
  cannot be explained by a nuclear disc.

\item In the \pmo{l}-\vgsr\ plane the nuclear disc stands out
as a continuous track of enhanced density.  The narrowness of the
track indicates a relatively low dispersion in the nuclear disc.

\item The asymmetry across $l=0\degrees$ and the non-zero slope
  $d\vgsr/d\pmo{l}$ at $\pmo{l} = 0\masyr$ are both signs that a
  nuclear disc is not axisymmetric.

\item A nuclear ring produces LOSVD peaks very similar to a nuclear
  disc; whether a nuclear disc or a nuclear ring is present can be
  determined by whether the track in the \pmo{l}-\vgsr\ plane is
  continuous or not.  Assuming that the high-\vgsr\ peaks observed by
  APOGEE at $l = 6\degrees$ are the tangent points of a nuclear
  structure, the ideal location to test whether a ring or a disc is
  present is at $l = 3\degrees$ to $4\degrees$.

\end{itemize}

\subsection{Distinction with the model of AS15}

Our model, and the model of AS15 are fundamentally very different so
it should be possible to tell them apart; in our model, an old disc or
ring of x2 orbits orthogonal to the bar gives rise to the
high-\vgsr\ peaks, while in the AS15 model, the high-\vgsr\ stars are
predominantly on x1 and higher order orbits, generally elongated
parallel to the bar, and which have recently been trapped by the bar.
In this model the high-\vgsr\ stars are preferentially young.  However
ages are always difficult to measure unambiguously so we turn to
kinematic differences between the models.  Because of the very
different orientations between the relevant orbits in the two models,
clear kinematic differences are expected.  The most promising
distinction between the two models is a geometric one that comes from
comparing the high-\vgsr\ stars at positive and negative longitudes.
At positive longitudes, APOGEE finds a statistically significant
secondary peak at $(l,b)=(6\degrees,0\degrees)$ \citep{zhou_shen+17}
with a velocity $\sim 220-250\kms$ (D15).  The x2 orbit model predicts
that at $\ = -6\degrees$ the peak velocity will appear at a lower
velocity, by $\sim 20\kms$.  In contrast, the model of AS15 predicts
that, in the absence of a dominant young stellar population, a
shoulder is present at a larger velocity, $\sim 250\kms$.  We propose
therefore that a very simple test of the two models can be produced by
comparing the mid-plane LOSVDs at $l = 6\degrees$ and $l =
-6\degrees$.  If the secondary peak is at lower $|\vgsr|$ in the $l =
-6\degrees$ field then this is evidence in favour of an x2 feature.
If instead the feature is at larger $|\vgsr|$, then this favours the
model of AS15.  In the absence of young stars at high-\vgsr, further
evidence in favour of the x2 model can be obtained if improved
statistics at the $(l,b) = (+6\degrees,0\degrees)$ field show that a
peak, rather than the shoulder predicted by the AS15 model, is
present.

\subsection{Conclusions}

We have presented predictions for the one-dimensional (LOSVD) and
two-dimensional (\pmo{l}-\vgsr) kinematics of a nuclear stellar ring
or disc.  Confirmation of such a system, which would be considerably
larger than the radius at which gas is now being delivered to the
Galactic center by the bar, would constitute an important clue to the
early evolution of the MW's bar.  APOGEE-2 will shortly be delivering
the LOSVD data at negative longitudes towards the bulge.  These data
have the potential to confirm or reject the presence of a
kiloparsec-scale nuclear x2-orbit structure.  Distinguishing whether
the structure is a ring or a disc requires proper motions and such
measurements of the required precision, while challenging, are already
possible \citep[e.g.][]{calamida+14}.  We therefore look forward to a
future possibility where the dynamical imprint of the early evolution
of the MW is captured in the fossil evidence at the centre.


\bigskip
\noindent
{\bf Acknowledgements.} 

\noindent
VPD is supported by STFC Consolidated grant \#~ST/M000877/1.  MN
acknowledges funding from the European Research Council under the
European Union's Seventh Framework Programme (FP 7) ERC Advanced Grant
Agreement n. 321035.  The simulation used in this paper was run at the
High Performance Computing Facility of the University of Central
Lancashire.  We thank Ralph Sch\"onrich, Monica Valluri and Juntai
Shen for fruitful discussion and Anita Kotla for proofreading the
paper.

\bigskip 
\noindent

\bibliographystyle{aj}
\bibliography{ms.bbl}

\label{lastpage}

\end{document}